\def\12{{1\over2}}
\def\bi{\bigskip}
\def\noi{\noindent}
\def\be{\begin{equation}}
\def\en{\end{equation}}
\def\bq{\begin{eqnarray}}
\def\eq{\end{eqnarray}}
\def\bc{\begin{center}}
\def\ec{\end{center}}
\def\beit{\begin{itemize}}
\def\eit{\end{itemize}}

\documentstyle[prl,aps,epsf]{revtex}
\begin{document}
\draft

\bigskip

\title{RESONANCE PROPAGATION AND THRESHOLD SINGULARITIES.}
\vskip2ex

\author{J.L. Lucio M.$^1$ \footnote{lucio@ifug.ugto.mx} and J. Pestieau $^2$.
\footnote{pestieau@fyma.ucl.ac.be} \vspace{.5cm} }

\address{$^1$ Instituto de F{\'\i}sica, Universidad de Guanajuato \\
Loma del Bosque \# 103, Lomas del Campestre, 37150 Le{\'o}n, Guanajuato; 
M{\'e}xico \\ \vspace{.3cm} $^2$Institut de Physique Th{\'e}orique, 
Universit{\'e} 
Catholique de Louvain \\ Chemin du Cyclotron 2, B-1348, Louvain La Neuve, 
Belgium.}

\maketitle

\vskip .5in

\begin{abstract}

\noi We consider the problem of propagation of an unstable particle in the 
framework of Quantum Field Theory. Using  unitarity, we show that a real 
renormalization constant free of threshold singularities naturally arises.

\end{abstract}


\bi

Comparison of the standard model of the electroweak interactions with high 
precision data involving the production of unstable particles ($W^\pm, 
Z^0$ and hopefully the Higgs in the near future) requires the incorporation 
of radiative corrections in the theoretical predictions. In particular, the 
particle's propagator is obtained via Dyson summation of the 
self-energy $A(s)$:

\be
P(s)~= ~{1 \over s-m^2_0- A(s)},            \label{prodys} 
\en

\noi where $m_0$ stands for the bare mass. In the conventional on-shell 
renormalization scheme (os) \cite{sirlin1}, the renormalized mass and width 
are defined as:

$$M^2~=~m_0^2~+~Re~A(M^2), \\ \eqno{(2a)}  $$
$$M\Gamma^{os}~=~-~{Im~A(M^2) \over 1~-~Re~A'(M^2)}, 
\eqno{(2b)}  \label{onshell}$$

\noi and the field renormalization constant is given by:

\bc
\be
\setcounter{equation}{3}
Z^{os}_2~=~{1 \over 1~-~Re~A'(M^2)}.   \label{zos}
\en
\ec
\noi The on-shell scheme provides a gauge invariant definition of the mass 
as long as the particle can be considered as stable. This is no longer true 
when the resonance width can not be neglected  \cite{sirlin2,stuart,valencia} 
and the formulation must be corrected in higher orders of perturbation theory 
by the addition of gauge dependent terms. It has been known for a long time 
that this problem can be solved by considering the mass $m$ and width 
$\Gamma$ of the unstable particle to be defined by the pole of the propagator 
\cite{sirlin2,stuart,valencia}. The position of the propagator's pole $s_p$
is obtained by solving the equation:

\be
s_p~=~m^2_0+A(s_p)~=~m^2-im\Gamma
\en

with

$$m^2~=~m^2_0~+~Re~A(s_p),\\   \eqno{(5a)}$$
$$m\Gamma~=~-Im~A(s_p).\eqno{(5b)} $$

\bi

Another problem of the on-shell scheme is that the renormalization constant 
suffers from threshold singularities. These singularities appear in 
$Re~A'(m^2)$ - but not in $Re~A(m^2)$, nor in $Im~A(m^2)$ - in the amplitude
describing the $S$ wave two body decays of scalar or vector resonances 
\cite{sirlin3,will}, when the mass of the decaying particle  approaches 
from below the mass threshold of the produced particles. Examples where this 
problem may be relevant have been discussed in \cite{sirlin3}. In fact we 
have to 
distinguish two different aspects: the first, is the relevance of the 
threshold singularities in the determination of the mass and width of the 
unstable particle, which is solved in the pole scheme \cite{will}, and second
the effect of the threshold singularities on the predictions of the theory 
for the production and decay rates. For the latter, a solution has been 
recently proposed, in the context of a gauge theory \cite{sirlin3}, by 
Kniehl,Palisoc and Sirlin (hereafter referred as KPS) 

\bi

 It is convenient at this point to recall the arguments 
used by KPS to find a formulation in which the threshold singularities 
are avoided. As far as we can see, the central points are the following:

\begin{itemize}
\item The pole position, and its interpretation in terms  of the physical mass 
and width of the unstable particle, lead to the relation, valid at the one 
loop level only:

\be
\setcounter{equation}{6}
Re~A'(m^2)~=~ {Im~A^{(1)}(m^2)~-~Im~A^{(1)}(s_p) \over m\Gamma}.
\label{identity}
\en

\noi The superscript refer to the number of quantum loops included in the 
computation of the self-energy. According to KPS, this serves as a regularized
version of $Re~A'(m^2)$, with the decay width $\Gamma$ serving as regulator.
\item In the {\bf pole scheme} the width of the resonance is defined by 
Eq.(5b). The KPS approach is based on the following identity:
\be
m\Gamma~=~-{Im~A(m^2) \over 1- {Im~A(m^2)-Im~A(s_p) \over  m\Gamma} }
\en
\noi Comparing with the {\bf on-shell} definition of width Eq.(2b),
KPS {\bf propose} that the regularized field renormalization constant should 
be given at all orders by:

\bi
\bq
{1 \over Z^{KPS}_2~}=~ 1- {Im~A(m^2)-Im~A(s_p) \over  m\Gamma}. 
\label{zsirlin} 
\eq
\end{itemize}

\noi Introducing Eq.(5b) into Eq.(9), we get

$${1 \over Z^{KPS}_2}~=~-{Im~A(m^2)  \over m\Gamma} \eqno{(8')} $$

\bi

\noi In this paper, we shall {\bf derive}, at all orders, Eq.(8').

\bi

Following Ref.\cite{sirlin4}, for {\bf real $s$}
we introduce the real and imaginary parts of the self-energy:

\bc\be
Re~A(s)~=~R(s),~~~~~~~~~~~~~~~~~~~Im~A(s)~=~I(s).  \label{aris}
\en\ec
\noi In terms of the pole position, the propagator, Eq.(\ref{prodys}) is 
expressed as:

\bc
\be
P(s)_{pole}~=~{1 \over F(s)}{1 \over s-s_p};        \label{pole}
\en
\ec
\noi where
\bc
\be
F(s)~=~ 1~-~{R(s)~-~R(s_p) \over s~-~s_p}~-~i~{I(s)~-~I(s_p) \over s~-~s_p}.
 \label{fs}
\en
\ec
\noi The field renormalization constant can be obtained from $F(s)$, 
usually after expanding around some value of $s~=~s_0$. Notice that if 
the point $s_0~=~s_p$ is chosen, a complex valued $Z^{-1}_2(s_p)
~=~1-A'(s_p)$ is obtained. The pole approach is based on the isolation of 
the pole, as in Eq.(\ref{pole}). We are not forced however to perform an 
expansion based on $s_p$ of the full 
quantity of interest (Green function or S matrix element). For example if in 
the present case such an 
expansion is carried for $F(s)$ we get a complex valued field renormalization 
constant, which most authors prefer to avoid. The common procedure is to 
perform an expansion of everything but the pole based on some real value of 
$s$.

\bi

Our starting point is the observation that, naively, one would expect in the 
pole approach  a complex field renormalization constant \cite{pesga} $Z^{-1}_2
(s_p)~=~1-A'(s_p)$, in contrast with the real $Z^{KPS}_2$ (Eq.(\ref{zsirlin})) 
found by KPS. We remark in this respect that KPS obtain $Z_2$ by comparing 
an identity following from the {\bf pole scheme}, with the  {\bf on-shell} 
definition of width where the $Z$ is real. Since the pole and on-shell 
schemes are equivalent through next to leading order, the procedure is fully
justified to that order.
\bi

It proofs convenient to consider

\be
{\sc T(s)}~=~{I(s) \over s-m^2_0-R(s)-iI(s)}.
\en
\bi

\noi {\sc T(s)} fulfills the unitarity relation: $Im~T(s)~=T(s)T^{\dagger}
(s),$ and $P(s)~=~{\sc T(s)}/I(s)$. The unitarity of $T(s)$ ensures that it 
can be expressed in the form:

\be
{\sc T(s)}~=~{-m\Gamma \over s~-m^2~-i~m\Gamma}e^{2i\delta (s)}~+~
{-y(s) \over 1~+~iy(s)},  \label{unita}
\en
where:
\be
e^{2i\delta (s)}~=~{1~-~iy(s) \over 1~+~iy(s)}
\en
\bi
If we want that  $P(s)~=~{\sc T(s)}/I(s)$, then $y(s)$ is given by:

\be
y(s)~=~{I(s)(s~-~m^2)~+~F(s)m\Gamma \over I(s)m\Gamma~-~F(s)(s-m^2)},
\en
with:
\be
F(s)~=~s-m^2_0-R(s)~=~s-m^2+Re~R(s_p)-R(s)-Im~I(s_p).
\en
\noi It is important to remark that $y(s)$ is real for real $s$. For our 
purposes it is better to express $T(s)$ not as in Eq.(\ref{unita}) but
in the equivalent form:
\be
{\sc T(s)}~=~{ -m\Gamma~-~(s-m^2)y(s) \over (s-s_p)(1~+~iy(s))},
\en
\noi as this allow us to conclude that:
\be
P(s)~=~{Z_2(s) \over (s-s_p)(1~+~iy(s))},
\en
where we have introduced:
\be
Z_2(s)~=~{-m\Gamma~-~(s-m^2)y(s) \over I(s)}   \label{zpole}
\en
\bi

This is our main result. We have renormalized the propagator in the pole 
scheme, using a real renormalization constant not involving $R'(m^2)$. The
following are the main characteristics of our procedure and result:

\begin{itemize}
\item The field renormalization we introduced in Eq.(\ref{zpole})
is free of threshold singularities as it only depends on $I(s)$ and $R(s)$.
\item We have made no assumption about the order of perturbation theory in 
which the self-energy has to be computed. In principle our result is valid 
to arbitrary order of perturbation theory.
\item For $s$ real, the field renormalization constant Eq.(\ref{zpole}) is
{\bf real}. Our result has the advantage that any point $s_0$ can be chosen to 
expand the Green function. A case of particular interest is $s_0=m^2$:
\be
Z_2(m^2)~=~-{m\Gamma \over I(m^2)}        \label{z2m}
\en

\noi This is precisely the $Z_2$ found by KPS (see Eq.(8') above and Eq.(23) 
in Ref.\cite{sirlin3}, and recall that $m\Gamma~=~-Im~A(s_p)$). Furthermore, 
when $s_0=m^2$ is chosen, we obtain for the propagator:
\be
P(s)~=~{Z_2(m^2) \over s-s_p}{1 \over 1~+~iy(m^2)} + ...  \label{p2m}
\en
\noi The ellipsis stand for terms of order $s-m^2$. The term  $1~+~iy(m^2)$ 
is not considered by KPS, this can be understood by noticing that $y(m^2)$ 
can be written as:
\bi

\be
y(m^2)~=~{Re~R(s_p)~-~R(m^2)~-Im~I(s_p) \over I(m^2)}  \\
~=~{m^2~-m^2_0~-~R(m^2) \over I(m^2)}   \label{equi}
\en

\noi The numerator of Eq.(\ref{equi}) is reminiscent of the on-shell mass 
definition Eq.(2a). In fact, $y(m^2)$ vanishes when the $m^2~=~
M^2$ equality holds, which is precisely the equivalence between the on-shell 
and pole schemes, which is valid only to leading order. Obviously the term 
$1~+~iy(m^2)$ is necessary to fulfill the unitarity requirement. 
\end{itemize} 

We can define a real field renormalization constant when $s_0=s_p$ (see the 
text beneath Eq.(\ref{fs})) by expanding $F(s)$:

\bq
F(s)~=~1-Re~A'(s_p)-iIm~A'(s_p)+ ...
~=~(1-Re~A'(s_p))(1-i{Im~A'(s_p)  \over 1-Re~A'(s_p)})+ ...
\eq

In this case, the ellipsis stand terms of order $(s-s_p)$. If we define:

\be
Z^{pole}_2 \equiv {1 \over 1-Re~A'(s_p)},          \label{z2p}
\en

we find

\be
P(s)_{pole}={Z^{pole}_2  \over s-s_p}{1 \over 1-i{Im~A'(s_p) \over  
1-Re~A'(s_p) }}.          \label{p2p}
\en

\noi To be compared with the on-shell field renormalization constant Eq.
(\ref{zos}) and $Z_2(m^2)$ Eq.(\ref{z2m}), while Eq.(\ref{p2p}) compares to
Eq.(\ref{p2m}). Notice that these expressions are equivalent only 
at the leading order.

\bi

In summary, we considered the problem of threshold singularities using the 
pole scheme to define the mass and width of the resonance. In this scheme, We 
introduced a {\bf real} field renormalization constant which is free of 
threshold singularities and proves the result obtained by Kniehl, Palisoc 
and Sirlin (see Eqs.(8) and (8')). The advantages of our approach are that 
{\it i)} It does not rely 
upon comparison of the conventional on-shell and pole schemes, {\it ii)} It  
is valid to arbitrary order of perturbation theory and {\it iii)} The 
field renormalization we introduce is a function of $s$, which can be
expanded around the value of $s$ better suited for each particular calculation.
\bi
\bi

\bc
ACKNOWLEDGMENTS
\ec

\bi
\bi
\noi J.L.L.M acknowledges financial support from CONACyT and hospitality from 
the Institut de Physique Th{\'e}orique, UCL, where part of this work was done.
\bi

\end{document}